\documentclass[aps]{article}

\begin{document}
\title{Neutrinos under Strong Magnetic Fields\footnote{%
Talk given at the Fourth Tropical Workshop on Particle Physics and
Cosmology, Cairns, Australia, June 9-13, 2003 }}
\author{Efrain J. Ferrer and Vivian de la Incera}
\date{Physics Department, State University of New York at Fredonia,
Fredonia, NY 14063, USA}
\maketitle
\begin{abstract}
In this talk we review the results on neutrino propagation under
external magnetic fields. We concentrate on the effects of strong
magnetic fields $M_{W}^{2}\gg B\gg m_{e}^{2}$ in neutral media. It
is shown that the neutrino energy density get a magnetic
contribution in the strong-field, one-loop approximation, which is
linear in the Fermi coupling constant as in the charged medium. It
is analyzed how this correction produces a significant oscillation
resonance between electron-neutrinos and the other two active
flavors, as well as with sterile neutrinos. The found resonant
level-crossing condition is highly anisotropic. Possible
cosmological applications are discussed. Effects due to primordial
hypermagnetic fields on neutrinos propagating in the symmetric
phase of the electroweak model are also presented. At sufficiently
strong hypermagnetic fields, $B\geq T^{2}$, the neutrino energy is
found to be similar to that of a massless charged particle with
one-degree of freedom.
\end{abstract}

\section{Introduction}

A wide spectrum of strong magnetic fields can be found and
predicted in nature. Magnetic fields can be as strong as
$10^{12}-10^{13}$ Gauss in the surface of typical radio pulsars
 \cite{Kronberg}, an even stronger ($10^{14}-10^{15}$ Gauss) in
magnetars (with interior fields that may range up to
$10^{16}-10^{17}$ Gauss) \cite{Duncan}. Superconducting
magnetic strings, if created after inflation, could generate fields of $%
10^{30}$ Gauss in their vicinity \cite{Witten}. Primordial
magnetic fields of $10^{24}$ Gauss at the electroweak (EW) scale
have been proposed as the possible origin of the seed field needed
to generate through galactic dynamo effect the large-scale
magnetic fields observed in a number of galaxies, and galaxy
clusters \cite{Grasso}.

Strong magnetic fields can affect matter since they confine
electrons perpendicular to its direction and consequently increase
the atom binding energies. This effect can create matter bound
states as molecular chains, magnetized three-dimensional condensed
matter, etc. \cite{Lai}. On the other hand, strong enough magnetic
fields have also important quantum electrodynamics effects, as the
modification of the dielectric property of the medium, the
polarization of photon modes \cite{Hugo} and the nonlinear photon
splitting \cite{Adler}.

In this talk we present the effect of strong magnetic fields on
the neutrino energy spectrum \cite{Efrain},\cite{Vivian} and
discuss its consequences for neutrino oscillations. Although the
neutrino, being a neutral particle, cannot directly interact with
the magnetic field, we know that its propagation can be modified
in the presence of an external field through quantum corrections. As shown below, in the presence of a strong magnetic field ($M_{W}^{2}\gg B\gg m_{e}^{2}$, where $M_{W}$ and $%
m_{e}$ are the W-boson and electron masses respectively) the
modification of the neutrino energy, due to one-loop corrections
of the self-energy operator at finite temperature, gives rise to a
resonant level-crossing condition in neutrino oscillations that is
linear in the Fermi coupling constant. As known, in the weak-field
approximation linear order modifications of the neutrino energy
only appears in charged media \cite{Nieves} (at zero field this is
the well known MSW effect \cite{msw}).

Taking into account that the particle-antiparticle asymmetry of
the universe is at the level of $10^{-10}-10^{-9}$, the
cosmological medium can be considered neutral, therefore for
primordial neutrino oscillations the MSW effect can be
disregarded. Nevertheless, if a strong primordial magnetic field
was present during the neutrino decoupling era, the results we
will discuss can be significant for cosmology, and specifically
for primordial nucleosynthesis.

In this talk we will also report some recent results
\cite{Cannellos} on neutrino propagation in a constant
hypermagnetic field. Effects of different nature in the presence
of primordial hypermagnetic fields have been also considered by
several authors \cite{Ayala}. If a primordial magnetic field
existed prior to the EW phase transition, only its U(1) gauge
component, i.e. the hypermagnetic field, would penetrate the EW
plasma for infinitely long times. Its non-Abelian component would
decay because of its infrared magnetic mass $\sim g^{2}T$, which
is generated non-perturbatively through the non-linear
interactions of the non-Abelian fields in the thermal bath
\cite{Linde}$.$ On the other hand, for Abelian and non-Abelian
electric fields a Debye screening will be always generated by
thermal effects producing a short-range decay for both fields
\cite{Fradkin}. Hence, the only large scale primordial field that
can penetrate the EW symmetric phase is the hypermagnetic field.

By studying the finite-temperature neutrino dispersion relations
in the chiral phase of the EW model in the presence of a strong
hypermagnetic field, we will show that the hypermagnetic field
counteracts the thermal effect responsible for the creation of
effective masses for the chiral leptons \cite{Weldon}$.$ As a
consequence, in the strong field approximation, where leptons are
basically constrained to the lower Landau level (LLL), we find
that neutrinos behave as massless particles with an anisotropic
propagation.

\section{Neutrino Self-Energy in Strong Magnetic Field}

To obtain the quantum correction to neutrino energy in a
magnetized thermal bath, we should calculate the neutrino
self-energy in the presence of a magnetic field at finite
temperature. As it is known, the neutrino self-energy operator is
a Lorentz scalar that can be formed in the spinorial space taking
the contractions with all the independent elements of the Dirac
ring. From its explicit chirality it reduces to

\begin{equation}
\sum (p)=R\overline{\sum }(p)L,\qquad \overline{\sum }(p)=V_{\mu
}\gamma ^{\mu }  \label{1}
\end{equation}
where $L,R=\frac{1}{2}(1\pm \gamma _{5})$ are the chiral-projector
operators, and $V_{\mu }$ is a Lorentz vector that in covariant
notation can be given as a superposition of four basic vectors
formed from the characteristic tensors of the corresponding
problem. In the present case
\begin{equation}
\overline{\sum }(p)=ap\llap/_{\| }+bp\llap/_{\perp }+cp^{\mu }%
\widehat{\widetilde{F}}_{\mu \nu }\gamma ^{\nu }+idp^{\mu
}\widehat{F}_{\mu \nu }\gamma ^{\nu }.  \label{2}
\end{equation}
The presence of the magnetic field, given through the
dimensionless magnetic
field tensor $\widehat{F}_{\mu \nu }$ and its dual $\widehat{\widetilde{F}}%
_{\mu \nu }$, allows the covariant separation in (\ref{2}) between
longitudinal and transverse momentum terms that naturally appears
in magnetic backgrounds

\begin{equation}
p\llap/_{\| }=p^{\mu }\widehat{\widetilde{F}}_{\mu \rho }%
\widehat{\widetilde{F}}_{\rho \nu }\gamma ^{\nu },\qquad
p\llap/_{\perp }=p^{\mu }\widehat{F}_{\mu \rho }\widehat{F}_{\rho
\nu }\gamma ^{\nu }. \label{3}
\end{equation}

The coefficients $a$, $b$, $c$, and $d$ are Lorentz scalars that
depend on the parameters of the theory and the approximation used.
We are interested in one-loop corrections, thus for a neutral
medium (i. e. in the absence of chemical potentials) the leading
contribution is given by the bubble diagram with internal lines of
virtual electrons and W-bosons. Since both virtual particles are
electrically charged, the magnetic field interacts with both of
them producing the Landau quantization of the corresponding
transverse
momenta. Thus, we end up with two set of Landau quantum numbers \cite{Efrain}%
, one for the electron, and other for the W-boson.

We assume a strong-field approximation, $M_{W}^{2}\gg B\gg
m_{e}^{2}$. Since in this case the gap between the electron Landau
levels is larger than the electron mass square, it is consistent
to use the LLL approximation for the electron, while for the
W-boson it is obvious that we must sum in all W-boson Landau
levels.

To justify such an approximation for cosmological applications we
should recall that due to the equipartition principle, the
magnetic energy can only be a small fraction of the universe
energy density. This argument leads to the relation between field
and temperature $B/T^{2}\sim 2$. For such fields,
the effective gap between the Landau levels (LL) is $eB/T^{2}\sim \mathcal{O}%
(1)$. In this case the weak-field approximation (where the sum in
all LL is important), cannot be used because field and temperature
are comparable. On the other hand, because the thermal energy is
of the same order of the energy gap between LL's, it is barely
enough to induce the occupation of just a few of the lower
electron LL's, since, as we are considering, the electron mass is
much smaller than the magnetic field. Therefore, it is natural to
expect that the LLL approximation in the electron spectrum will
provide a good qualitative description of the neutrino propagation
in the presence of strong magnetic fields, although clearly a more
quantitative treatment of the problem in a field $B\sim 2T^{2} $
would require numerical calculations due to the lack of a leading
parameter.

In the leading order in the expansion in powers of the Fermi
coupling constant the scalar coefficients of Eq. (\ref{2}) are
obtained in our approximation as \cite{Efrain},\cite {Vivian}

\begin{equation}
a=-c\simeq \frac{g^{2}eB}{8\pi M_{W}^{2}}[\frac{1}{4\pi }-\frac{T^{2}}{%
3M_{W}^{2}}]\exp (-p_{\perp }^{2}/eB),\qquad b=d\simeq 0
\label{4}
\end{equation}
Notice that by comparison the thermal contribution is smaller in a factor of $%
1/M_{W}^{2}$ with respect to the field-dependent vacuum
contribution. Then, in the strong-field limit the thermal
contribution has the same second order in the Fermi coupling
constant as it is in the zero- \cite{Raffelt}, and weak-field
\cite{Elmfors} cases. On the other hand, the self-energy
field-dependent vacuum contribution in (\ref{4}) is of the same
order in the
Fermi coupling constant as the one found in a charged medium at zero %
 \cite{msw} and weak \cite{Nieves} fields.

Using the zero-temperature weak-field results of Ref.
\cite{McKeon} to identify the scalar coefficients of the general
structure (\ref {2}) in that approximation, we have that in the
weak-field limit they are given by

\begin{equation}
a=b\simeq 0,\qquad c=-\frac{6}{4}d\simeq \frac{6g^{2}eB}{(4\pi )^{2}M_{W}^{2}%
}  \label{5}
\end{equation}

We see that at weak field, the neutrino self-energy has also
linear contributions in the Fermi coupling constant, but they are
associated with different structures in (\ref{2}) as compared with
the result in the strong-field limit (\ref{4}). The role of the
different self-energy structural members into the neutrino energy
spectrum will be clear in the next section. There, we will show
that the results in the strong-field limit (\ref{4}) produce a
magnetic field dependence in the neutrino energy which is linear
in the Fermi coupling constant, while the weak-field results
(\ref{5}) produce a smaller second-order contribution.

\section{Neutrino Energy Spectrum and Index of Refraction}

The neutrino field equation of motion is

\begin{equation}
\lbrack p\llap/-\sum (p)]\Psi _{L}=0  \label{6}
\end{equation}
The dispersion relation is obtained by solving Eq. (\ref{6}), or
equivalently, by finding the nontrivial solution of Eq. (\ref{6})
through the equation

\begin{equation}
\det [p\llap/-\sum (p)]=0  \label{7}
\end{equation}
where $\sum (p)$ is given in its general covariant form by Eq.
(\ref{2}).

In the strong-field limit (\ref{4}), the solution of Eq. (\ref{7})
is \cite{Efrain}

\begin{equation}
E_{p}=\pm \left|
\overrightarrow{p}+\sqrt{2a}(\overrightarrow{p}\times
\widehat{\overrightarrow{B}})\right| =\pm \left|
\overrightarrow{p}\right| [1+2a\sin ^{2}\alpha ]  \label{8}
\end{equation}
where $\alpha $ is the angle between the direction of the neutrino
momentum and that of the applied magnetic field.

To obtain the neutrino index of refraction $n$, we substitute
(\ref{8}) into the formula

\begin{equation}
n\equiv \frac{\left| \overrightarrow{p}\right| }{E_{p}}  \label{9}
\end{equation}
to find \cite{Efrain}

\begin{equation}
n\simeq 1-a\sin ^{2}\alpha  \label{10}
\end{equation}

From Eqs. (\ref{8}) and (\ref{10}) we can see that neutrinos
moving with different directions in the magnetized space will have
different dispersion relations and consequently, different indexes
of refraction. It is interesting to notice that, although
neutrinos are electrically neutral particles, the magnetic field,
through quantum corrections, can produce anisotropic neutrino
propagation, having a maximum effect for neutrinos propagating
perpendicularly to the magnetic field (see Eq. (\ref{10})). The
order of the energy correction is $g^{2}\frac{\left| eB\right|
}{M_{W}^{2}}$, and so is the order of the asymmetry.

If we consider the weak-field results (\ref{5}) in the dispersion relation (%
\ref{7}) we obtain

\begin{equation}
E_{p}^{\prime }=\pm \left| \overrightarrow{p}\right| [1+\frac{5}{18}%
c^{2}\sin ^{2}\alpha ]  \label{11}
\end{equation}
Here, as the energy depends on $c^{2}$, we can see that the weak
field produces a negligible second order in the Fermi coupling
constant expansion. It can be corroborated that the inclusion of
temperature in this approximation also produces a second order
correction \cite{Elmfors}.

An asymmetric neutrino propagation of linear order in the Fermi
coupling constant expansion was previously found \cite{Nieves} for
weak magnetic fields but in a charged medium ($\mu \neq 0$).
There, the energy correction was given by

\begin{equation}
E_{p}^{\prime \prime }=a^{\prime }\pm \left| \overrightarrow{p}-b^{\prime }%
\widehat{\overrightarrow{B}}\right| ,  \label{12}
\end{equation}
where the coefficients $a^{\prime }$ and $b^{\prime }$ are
proportional to the electron/positron number densities $n_{\pm }$,
and electron/positron distribution functions  $%
f_{\pm }$ respectively

\begin{equation}
a^{\prime }=\frac{g^{2}}{4M_{W}^{2}}(n_{-}-n_{+}),\qquad b^{\prime }=\frac{%
eg^{2}}{2M_{W}^{2}}\int \frac{d^{3}p}{(2\pi
)^{3}2E}\frac{d}{dE}(f_{-}-f_{+}) \label{13}
\end{equation}

If we compare our results for strong magnetic fields in a neutral medium (%
\ref{8}) with those for weak field in a charged medium (\ref{12}),
we see that while in the first case the maximum field effects
occur for neutrinos propagating perpendicularly to the field
direction, in the second case this propagation modes are not
affected by the field, but on the contrary, the maximum effect
takes place for neutrinos propagating along the field lines.
Moreover, the asymmetric term in the dispersion relation found in
Ref. \cite {Nieves} changes its sign when the neutrino reverses
its motion. This property is crucial for a possible explanation of
the peculiar high pulsar velocities \cite{Kusenko}. Nevertheless,
in (\ref{8}) we find that in our approximation the neutrino
energy-momentum relation is invariant under the change of $\alpha
$ by $-\alpha $. Finally, the dispersion relations (\ref {12})
have different values for neutrinos and antineutrinos
respectively,
since in a charged medium the CP-symmetry is violated. In our result (\ref{8}%
), particle and antiparticle have the same energy, as it should be
in a neutral medium.

\section{Neutrino Oscillations and Resonance in  a Strongly Magnetized Neutral Medium}

Assuming that the magnetic field strength is confined within the
range $m_{e}^{2}\ll B\ll m_{\mu }^{2}$, the strong-field
approximation becomes valid for electron-neutrinos, but not for
muon-neutrinos and tau-neutrinos. For the last two, the magnetic
field will have the weak effect in the energy corrections
discussed below Eq. (\ref{11}) and therefore can be neglected.

Let us consider the evolution equation in the presence of a strong
magnetic field for a two-level system

\begin{equation}
\frac{d}{dt}\left(
\begin{array}{l}
\nu _{e} \\
\nu _{\mu }
\end{array}
\right) =H_{B}\left(
\begin{array}{l}
\nu _{e} \\
\nu _{\mu }
\end{array}
\right)  \label{14}
\end{equation}
where the Hamiltonian $H_{B}$ is given by

\begin{equation}
H_{B}=p+\frac{m_{1}^{2}+m_{2}^{2}}{4p}+\left(
\begin{array}{ll}
-\frac{\Delta m^{2}}{4p}\cos 2\theta +E_{B} & \frac{\Delta
m^{2}}{4p}\sin
2\theta  \\
\frac{\Delta m^{2}}{4p}\sin 2\theta  & \frac{\Delta m^{2}}{4p}\cos
2\theta
\end{array}
\right)   \label{15}
\end{equation}
Here, $\theta $ is the  vacuum mixing angle, $\Delta
m^{2}=m_{2}^{2}-m_{1}^{2}$ is the mass square difference of the
two mass eigenstates, and the magnetic energy density contribution
to the electron-neutrino is

\begin{equation}
E_{B}=\frac{g^{2}eB}{2(4\pi )^{2}M_{W}^{2}}\left|
\overrightarrow{p}\right| \sin ^{2}\alpha  \label{16}
\end{equation}
In Eq. (\ref{15}) the magnetic field contribution to the
muon-neutrino in the second diagonal term has been neglected,
taking into account that it will be of second order in the Fermi
coupling constant as corresponds to the weak-field approximation.

To find the evolution Hamiltonian corresponding to the mass
eigenstates in the magnetized space we need to transform $H_{B}$
according to

\begin{equation}
\widetilde{H}=U_{B}^{t}H_{B}U_{B}  \label{17}
\end{equation}
with transformation matrix

\begin{equation}
U_{B}=\left(
\begin{array}{ll}
\cos 2\theta _{B} & \sin 2\theta _{B} \\
-\sin 2\theta _{B} & \cos 2\theta _{B}
\end{array}
\right)  \label{18}
\end{equation}
depending on the new mixing angle $\theta _{B}$ given through the
relation

\begin{equation}
\tan 2\theta _{B}=\frac{\tan 2\theta }{\frac{\Delta m^{2}}{2p}\cos
2\theta -E_{B}}  \label{19}
\end{equation}

If we consider that the only flavor present at the initial time
was the electron-neutrino, using the evolution equation (\ref{15})
we find that the appearance probability for the muon neutrino is
given by

\begin{equation}
P_{B}\left( \nu _{e}\rightarrow \nu _{\mu }\right) =\sin
^{2}\theta _{B}\sin ^{2}\frac{\pi x}{\lambda }  \label{20}
\end{equation}
where $\lambda $ is the oscillation length in the magnetized space

\begin{equation}
\lambda =\frac{\lambda _{0}}{[\sin ^{2}2\theta +(\cos 2\theta
-\frac{\lambda _{0}}{\lambda _{e}})]^{1/2}}  \label{21}
\end{equation}
written in terms of the vacuum ($\lambda _{0}$) and magnetic
($\lambda _{e}$) oscillation lengths

\begin{equation}
\lambda _{0}=\frac{4\pi p}{\Delta m^{2}},\qquad \lambda _{e}=\frac{2\pi }{%
E_{B}}  \label{22}
\end{equation}
and

\begin{equation}
\sin ^{2}2\theta _{B}=\frac{\sin ^{2}2\theta }{(\cos 2\theta
-\frac{\lambda _{0}}{\lambda _{e}})^{2}+\sin ^{2}2\theta }
\label{23}
\end{equation}is the probability amplitude defined through the mixing angle
$\theta _{B}$ (\ref{19}).

If the resonant condition

\begin{equation}
\frac{\lambda _{0}}{\lambda _{e}}=\cos 2\theta   \label{24}
\end{equation}
is satisfied, then the probability amplitude (\ref{23}) will get
is maximum value independently of the value of the mixing angle in
vacuum $\theta $. The condition (\ref{24}) is a resonant
level-crossing condition, and as usual, it can be also obtained by
equating the two diagonal Hamiltonian elements in (\ref{15}). At a
magnetic field for which the condition (\ref{24}) is satisfied, a
maximum transmutation between the two flavors will occur (the same
effect will be obtained between electron-neutrinos and tau or
sterile neutrinos). We should notice that the resonant effect in
the magnetized neutral medium will be anisotropic, depending on
the direction of propagation of the electron-neutrino with
respect to the magnetic field (see that $\lambda _{e}$ depends on $\alpha $%
).

The resonant phenomenon here is similar to that in the well known
MSW effect in a charged medium. Nevertheless, we should stress
that in the magnetized neutral medium the oscillation process will
not differentiate between neutrinos and antineutrinos, while in
the charged medium, we have that if there is resonance for the
neutrino there will be no resonance for the antineutrino and
viceversa.

\section{Consequences for Cosmology}

The early Universe, unlike the dense stellar medium, is almost
charge symmetric ($\mu =0$), since, as already mentioned, the
particle-antiparticle asymmetry in the Universe is believed to be
at the level of $10^{-10}-10^{-9}$, while in stellar material it
is of order one. It is known that the contribution to the neutrino
energy density of pure thermal effects \cite{Raffelt}, or of
quantum corrections obtained in a weakly magnetized neutral medium
\cite{Elmfors},\cite {McKeon}, are both of second order in the
Fermi coupling constant, therefore negligible small. Nevertheless,
as we have shown in (\ref{8}), if sufficiently strong magnetic
fields were present in the early Universe, they would give rise to
corrections to the energy density that are linear in the Fermi
coupling constant. These corrections can produce effects as
significant as those associated to the MSW mechanism.

The existence of strong magnetic fields in the early Universe
seems to be a very plausible idea \cite{Grasso},\cite{enq}$,$ as
they may be required to explain the observed galactic magnetic
fields, $B\sim 2\times 10^{-6}$ $G$ on scales of the order of
$100$ $kpc$ \cite{Kronberg},\cite{Galaxies}$.$

The strength of the primordial magnetic field in the neutrino
decoupling era can be estimated from the constraints derived from
the successful nucleosynthesis prediction of primordial $^{4}He$
abundance \cite{He}$,$ as well as on the neutrino mass and
oscillation limits \cite{Enqvist}. These constraints, together
with the energy equipartition principle, lead to the relations

\begin{equation}
m_{e}^{2}\leq eB\leq m_{\mu }^{2},\qquad B/T^{2}\sim 2 \label{25}
\end{equation}
Then, it is reasonable to assume that between the QCD phase
transition epoch and the end of nucleosynthesis a primordial
magnetic field in the range given by relations (\ref{25}) could
have been present \cite{Efrain}.

If such a field existed, it could significantly modify the $\nu
_{e}\leftrightarrow \nu _{\mu },\nu _{\tau }$ and $\nu
_{e}\leftrightarrow \nu _{s}$ resonant oscillations in the way we
have shown in this paper, and consequently affect primordial
nucleosynthesis \cite {Dolgov}.

Another interesting question regarding the effect of strong
primordial fields is how these fields would affect neutrino
propagation prior to the EW phase transition in case they were
originated at earlier times in the Universe evolution. Since in
this phase only a hypermagnetic fields matters, as the non-Abelian
component decays due to the acquired infrared mass, it is enough
to investigate the propagation of neutrinos in the presence of a
background hypermagnetic field.

Notice that there are essential differences between the
interactions of neutrinos with hypermagnetic and magnetic fields.
We recall that the neutrino, being a neutral particle with respect
to the electromagnetic group, has instead non-zero hypercharge and
thus can interact with a hypermagnetic field already at the bare
level.

In the symmetric phase of the EW model, fermions are in the
chirally invariant phase with separated left-handed and
right-handed representations of the gauge group. In that phase, a
fermion mass term in the Lagrangian density is forbidden by the
symmetry of the theory. However, as it was found in Ref.
\cite{Weldon}$,$ finite temperature corrections induce a pole in
the fermion Green's function that plays the role of an effective
mass. The induced effective ''mass'' modifies the fermion
dispersion relation in the primeval plasma opening the door for
possible cosmological consequences
\cite{Weldon}$,$\cite{Dolgov}$.$

As it was shown in \cite{Cannellos} a strong hypermagnetic field
can modify the neutrino self-energy at finite temperature, so at
the one-loop level it is given by

\begin{equation}
\Sigma _{\nu _{L}}(\overline{p})=\left[ Ap\llap/_{\| }+B\,p^{\mu
}\widehat{\widetilde{H}}_{\mu \nu }\gamma ^{\nu }\right] L
\label{26}
\end{equation}
with

\begin{equation}
A=-B=\frac{G_{\nu _{L}}^{2}}{2\pi }\left[ \frac{a}{4\pi }+\frac{T^{2}}{%
\left| g^{\prime }H\right| }\right]   \label{27}
\end{equation}
In (\ref{27}), $G_{\nu _{L}}^{2}=[\frac{(g^{\prime
})^{2}}{4}+\frac{3(g)^{2}}{4}]$, $a$
is a positive constant of order one ($a\simeq 0.855$), and $\widehat{%
\widetilde{H}}_{\mu \nu }$ is the dual of the dimensionless
hypermagnetic field tensor ($\widehat{H}_{\mu \nu }=H_{\mu \nu
}/H$).

The neutrino dispersion relation in the presence of a constant
hypermagnetic field including radiative corrections is given by

\begin{equation}
\det \left[ \overline{p}\cdot \gamma +\Sigma \right] =0,
\label{28}
\end{equation}
where the neutrino generalized momentum in the presence of the
hypermagnetic field is $\overline{p}_{\mu
}=(p_{0},0,-sgn(g^{\prime }H)\sqrt{\left| g^{\prime }H\right|
n},p_{3})$, with the integer $n=0,1,2,...$ labelling the neutrino
Landau levels.

The fact that the two structures in (\ref{26}) have the same
coefficient (i.e. $A=-B$), together with the dependence of
$\overline{p}_{\mu }$ in the strong-field approximation on its
longitudinal components $p_{0}$ and $p_{3}$ only (i.e. in the LLL
approximation ($n=0$)), has significant consequences for the
propagation of chiral leptons in the hypermagnetized medium, as
one can immediately corroborate by explicitly solving the lepton
dispersion relation (\ref{28}) to obtain

\begin{equation}
-(1+2A)p_{0}^{2}+(1+2A)p_{3}^{2}=0  \label{29}
\end{equation}
This result indicates that in a strong hypermagnetic field
neutrinos behave as massless particles, so the hypermagnetic field
counteracts the temperature effects on the neutrino dispersion
relation. As mentioned above, according to Ref. \cite{Weldon} the
dispersion relations of the chiral leptons are modified due to
high-temperature effects in such a way that a
temperature-dependent pole appears in their one-loop Green's
functions (i.e. an effective mass). As we have shown here, if a
primordial hypermagnetic field could exist in the EW plasma prior
to the symmetry-breaking phase transition, with such a strength
that the leptons would be mainly confined to the LLL for the
existing temperatures, then the thermal effective mass found for
the leptons in that phase at zero field \cite{Weldon} will be
swept away. This effect can be of interest for cosmology, since it
will alter the behavior of the lepton masses with temperature
during the EW phase transition.

Another important outcome of the dispersion relation (\ref{29}) is
the large anisotropy of the neutrino propagation in strong
hypermagnetic field domains. This is the consequence of the
degeneracy in the energy, which does not depend on the tranverse
momentum components. No degeneracy appears in the case of
neutrinos propagating in magnetic fields (see Eq. (\ref{8})). This
different behavior can be understood from the fact that the
neutrino is minimally coupled to the hypermagnetic field because
of its nonzero hypercharge, but due to its electrical neutrality,
it can couple to the magnetic field only through radiative
corrections.

It is worth to notice that the leading contribution to the
one-loop correction of the neutrino self-energy at small momentum
gives rise to a dispersion relation that is formally the same as
the bare dispersion relation in the strong hypermagnetic field (it
is the typical dispersion relation of a massles charged particle
in a magnetic field strong enough to confine the particle to its
LLL).

From the above results, one can envision that if large primordial
fields were present in the early universe, the neutrino
propagation would be highly anisotropic, an effect that could
leave a footprint in a yet to be detected relic neutrino cosmic
background. If such a footprint were observed, it would provide a
direct experimental proof for the existence of strong primordial
fields.

\section{Concluding Remarks}

In this talk we have studied, within the framework of the standard
model, the quantum effects on neutrinos propagating in a strong
magnetic field. For fields in the range $m_{\mu}^{2}\gg B\gg
m_{e}^{2}$, we found a field-dependent contribution to the
electron-neutrino energy density that is linear in the Fermi
coupling constant. For the other neutrino flavors, the
strong-field approximation is not valid and consequently the
magnetic corrections to the energy density are negligible small.

This separation between the magnetic contribution to the energy
density of different flavors directly affects neutrino
oscillations leading to a level-crossing resonance that depends on
the value of the applied magnetic field. This resonant effect is
similar to the well known MSW effect, but with the difference that
it can take place in a neutral medium, and that it has an
anisotropic character, depending on the direction of the neutrino
propagation with respect to the magnetic field.

The fact that a resonant oscillation can take place in a neutral
medium at the leading order in the Fermi coupling constant has
significant interest for cosmology in case that a sufficiently
strong primordial field were present during the neutrino
decoupling era, since it would affect primordial nucleosynthesis.

On the other hand, in the symmetric phase of the EW theory the
effect of a primordial field can only be carried out by the
hypermagnetic field. In the presence of this field, neutrinos
behave as charged particles, and in the strong field
approximation, when leptons are confined to their LLL, the leading
contribution to the energy density of the neutrino thermal bath is
similar to that of a massless charged particle with only one
degree of freedom (the energy only depends on the longitudinal
momentum). That is, the strong hypermagnetic field swept away the
thermal mass found at zero field and high temperature
\cite{Weldon}. This result can be important for the study of the
EW phase transition at finite temperature and in the presence of
strong primordial fields.

\section{Acknowledgment}

This work was supported in part by the National Science Foundation
under Grant No. PHY-0070986.


\begin{thebibliography}{99}
\bibitem{Kronberg}  P. P. Kronberg, Rep. Prog. Phys., \textbf{57}, 325 (1994); R. Beck et. al., Ann. Rev. Astron. Astrophys. \textbf{34}, 153 (1996).

\bibitem{Duncan}  R. C. Duncan and C. Thompson, Astrophys. J.
\textbf{392}, L9 (1992); B. Paczynski, Acta Astron. \textbf{42},
145 (1992); V. V. Usov, Nature \textbf{357}, 472 (1992); C.
Thompson and R. C. Duncan Astrophys. J. \textbf{473}, 322
 (1996).

\bibitem{Witten}  J. P. Ostriker, C. Thompson and E. Witten, Phys. Lett. B
\textbf{180},  231 (1986).

\bibitem{Grasso}  D. Grasso and H.R. Rubinstein, Phys. Rep.
\textbf{348}, 163 (2001).

\bibitem{Lai}  For a recent review see D. Lai, Rev. Mod. Phys.
\textbf{73}, 629 (2001).

\bibitem{Hugo}  H. Perez-Rojas ans A. E. Shabad, Ann. of Phys.
\textbf{121}, 432 (1979); \textbf{138}, 1 (1982); H. Perez-Rojas,
Zh. Eksp. Teor. Fiz. \textbf{76}, 1 (1979); A. K. Ganguly, S.
Konar and P. B. Pal, Phys. Rev. D \textbf{60}, 105014 (1999); J.
C. D'Olivo, J. F. Nieves and S. Sahu, Phys. Rev. D \textbf{67},
025018 (2003).

\bibitem{Adler}  S. L. Adler, Ann. Phys. \textbf{67}, 599 (1971); V. O.
Papanyan and V. I. Ritus, Zh. Eksp. Teor. Fiz. \textbf{61}, 2231
(1971) (Sov. Phys. JETP \textbf{34}, 1195 (1972)); \textit{ibid }
\textbf{65}, 1756 (1973), (Sov. Phys. JETP \textbf{38}, 879
(1974)); V. N. Baier, A. I. Milstein, and R. Zh. Shaisultanov,
Phys. Rev. Lett. \textbf{77}, 1691 (1996); J. I. Weise, M. G.
Baring, and D. B. Melrose Phys. Rev D\textbf{57}, 5526 (1998); A.
V.
Kuznetsov, N. V. Mikheev, and M. V. Chistyakov, Phys. of Atom. Nuc. \textbf{%
62}, 1638 (1999).

\bibitem{Efrain}  E. Elizalde, E. J. Ferrer and V. de la Incera, Ann. of
Phys. \textbf{295}, 33 (2002); E, J. Ferrer, "Non-Perturbative
Effects of Strong Magnetic Fields" in Quantization, Gauge Theory,
and Strings, edited by A. Semikhatov, M. Vasiliev and V. Zaikin,
Proc. of the International Conference dedicated to the memory of
Prof. Efim Fradkin, Sci. World Pub. Co. 2001, pp. 301-306.

\bibitem{Vivian}  E. Elizalde, E. J. Ferrer and V. de la Incera, \textit{%
Neutrino Propagation in a Strongly Magnetized Medium}, in
preparation.

\bibitem{Nieves}  J. C. D'Olivo, J. F. Nieves and P. B. Pal, Phys. Rev. D%
\textbf{40}, 3679 (1989).

\bibitem{msw}  L. Wolfenstein, Phys. Rev. D\textbf{17}, 2369 (1978); S. P.
Mkheyev, A. Yu. Smirnov, Sov. J. Nucl. Phys. \textbf{42}, 913
(1985).

\bibitem{Cannellos}  J. Cannellos, E. J. Ferrer and V. de la Incera, Phys.
Lett. B\textbf{542} 123 (2002).

\bibitem{Ayala} M. Giovannini and M. E. Shaposhnikov, Phys. Rev. D\textbf{57}, 2186 (1998); M. Giovannini
Phys. Rev. D\textbf{61}, 063004 (2000); \textit{ibid } 063502; A.
Ayala, J. Besprosvany, G. Pallares and G. Piccinelli, Phys. Rev.
D\textbf{64}, 123529 (2001); A. Ayala, G. Piccinelli and G.
Pallares, Phys. Rev. D\textbf{66}, 1103503 (2002); A. Ayala, and
J. Besprosvany, Nucl. Phys. B\textbf{651}, 211 (2003).

\bibitem{Linde}  A.D. Linde, Rep. Prog. Phys., \textbf{42}, 389 (1979); Phys.
Lett B\textbf{96}, 178 (1980); K. Kajantie, M. Laine, K.
Rummukainen, and M. Shaposhnikov, Nucl. Phys. B\textbf{493}, 413
(1997).

\bibitem{Fradkin}  E.S. Fradkin, Proceedings Lebedev Phys. Inst., Vol.
\textbf{29}, 7 (1965) (Eng. Transl., Consultant Bureau, New York
(1967)); A. K. Rebhan, Phys. Rev. D\textbf{48}, R3967 (1993);
Nucl. Phys. B\textbf{430}, 319 (1994); E. Braaten, and A. Nieto,
Phys. Rev. Lett \textbf{73}, 2402 (1994); R. Baier, and O.K.
Kalashnikov, Phys. Lett B\textbf{328}, 450 (1994).

\bibitem{Weldon}  H.A. Weldon, Phys. Rev. D\textbf{26}, 2789 (1982).

\bibitem{Raffelt}  D. Notzold and G. Raffelt, Nucl. Phys.
B\textbf{307}, 924 (1988).

\bibitem{Elmfors}  P. Elmfors, D. Grasso, G. Rafflet, Nucl. Phys. B\textbf{%
479}, 3 (1996).

\bibitem{McKeon}  G. McKeon, Phys. Rev. D\textbf{24}, 2744 (1981).

\bibitem{Kusenko}  A. Kusenko and G. Segre, Phys. Rev. Lett.
\textbf{77}, 4872 (1996).

\bibitem{enq}  K. Enquist, Int. J. Mod. Phys. D\textbf{7}, 331 (1998).

\bibitem{Galaxies}  Y. Sofue, M. Fujimoto, and R. Wielebinski, Ann. Rev.
Astron. Astrophys. \textbf{24}, 459 (1986); R. Beck et. al., Ann.
Rev. Astron. Astrophys. \textbf{34}, 153 (1996).

\bibitem{He}  K. A. Olive, D. N. Schramm, G. Steigman, and T. Walker, Phys.
Lett. B\textbf{\ 236}, 454 (1990); L. M. Krauss and P. Romanelli,
Astrophys. J. \textbf{358}, 47 (1990).

\bibitem{Enqvist}  K. Enqvist, V. Semikoz, A. Shukurov, and D. Sokoloff,
Phys. Rev. D\textbf{48}, 4557 (1993).

\bibitem{Dolgov}  A. D. Dolgov, Phys. Rep. \textbf{370}, 333 (2002).


\end{thebibliography}
\end{document}